%Paper: hep-ph/9308317
%From: butler@phy-server.Phy.QueensU.CA (Malcolm Butler)
%Date: Sun, 22 Aug 93 15:48:33 EDT

\input harvmac
%%%%%%%%%%%%%%%%%%%%%%%%%%%%%%%%%%%%%%%%%%%%%%%%%%%%%%%%%%%%%%%%%%%%%
%
%  UCSD macros to overwrite some of the definitions in harvmac.tex
%  (include after harvmac.tex)
%  last modified 4/92
%
%%%%%%%%%%%%%%%%%%%%%%%%%%%%%%%%%%%%%%%%%%%%%%%%%%%%%%%%%%%%%%%%%%%%%

%%
%
% modify the output routine for the little format
%
\ifx\answ\bigans
\else
\output={
\almostshipout{\leftline{\vbox{\pagebody\makefootline}}}\advancepageno}
\fi
%
%
% address
%
\def\mayer{\vbox{\sl\centerline{Department of Physics 0319}%
\centerline{University of California, San Diego}
\centerline{9500 Gilman Drive}
\centerline{La Jolla, CA 92093-0319}}}
%
% grant numbers
%

%
% preprint number
%
% restores pagenumbers
%
% abstract
%
\def\abstract#1{\centerline{\bf Abstract}
\nobreak\medskip\nobreak\par  #1}
%
%
% titlefont
%
%
\edef\tfontsize{ scaled\magstep3}
 \tfontsize  \tfontsize
 \tfontsize \font\titlei=cmmi10 \tfontsize
\font\titleis=cmmi7 \tfontsize \font\titleiss=cmmi5 \tfontsize
\font\titlesy=cmsy10 \tfontsize \font\titlesys=cmsy7 \tfontsize
\font\titlesyss=cmsy5 \tfontsize  \tfontsize
\skewchar\titlei='177 \skewchar\titleis='177 \skewchar\titleiss='177
\skewchar\titlesy='60 \skewchar\titlesys='60 \skewchar\titlesyss='60
%
%\def\titlefont{\def\rm{\fam0\titlerm}% switch to title font
%\textfont0=\titlerm \scriptfont0=\titlerms

%\scriptscriptfont0=\titlermss
%\textfont1=\titlei \scriptfont1=\titleis

%\scriptscriptfont1=\titleiss
%\textfont2=\titlesy \scriptfont2=\titlesys

%\scriptscriptfont2=\titlesyss
%\textfont\itfam=\titleit \def\it{\fam\itfam\titleit}\rm}
%
%
% math symbols
%
%--------------------------------------------------------------------

%
\def\inv{^{\raise.15ex\hbox{${\scriptscriptstyle -}$}\kern-.05em 1}}
  %prime
\def\lbar{{\lower.35ex\hbox{$\mathchar'26$}\mkern-10mu\lambda}}

%lambda bar

%
%
% various slashed symbols
%
%
 % slashes a character
\def\dsl{\,\raise.15ex\hbox{/}\mkern-13.5mu D}

%this one can be subscripted
\def\delsl{\raise.15ex\hbox{/}\kern-.57em\partial}
\def\Ksl{\hbox{/\kern-.6000em\rm K}}
\def\Asl{\hbox{/\kern-.6500em \rm A}}
\def\Dsl{\hbox{/\kern-.6000em\rm D}} %roman D
\def\Qsl{\hbox{/\kern-.6000em\rm Q}}
\def\gradsl{\hbox{/\kern-.6500em$\nabla$}}
%
% space and backspace in l mode
%
\def\lspace{\ifx\answ\bigans{}\else\qquad\fi}
\def\lbspace{\ifx\answ\bigans{}\else\hskip-.2in\fi} % $$\lbspace...$$
%
%     boxes an equation
%
\def\boxeqn#1{\vcenter{\vbox{\hrule\hbox{\vrule\kern3pt\vbox{\kern3pt
        \hbox{${\displaystyle #1}$}\kern3pt}\kern3pt\vrule}\hrule}}}
%
%     draw a little box (end of proof symbol)
%     e.g. \mbox{.1}{.1}
%
\def\mbox#1#2{\vcenter{\hrule \hbox{\vrule height#2in
\kern#1in \vrule} \hrule}}
%
%
%
%     curly letters
%
   %curly letters
 
\def\CC{{\cal C}} \def\CD{{\cal D}}
 \def\CF{{\cal F}}
\def\CG{{\cal G}} \def\CH{{\cal H}}
 
 \def\CL{{\cal L}}
\def\CM{{\cal M}}

%
%
%
%     derivatives
%
%

%

%
%

\def\darr#1{\raise1.5ex\hbox{$\leftrightarrow$}\mkern-16.5mu #1}

%
 %pound sterling
%
 %puts a small half in a displayed

\def\frac#1#2{{\textstyle{#1\over #2}}} %puts a small fraction

%in a displayed eqn
%
%
%     various math operators
%
%

%
%
%
%
%
%       relations
%
\def\ltap{\ \raise.3ex\hbox{$<$\kern-.75em\lower1ex\hbox{$\sim$}}\ }
\def\gtap{\ \raise.3ex\hbox{$>$\kern-.75em\lower1ex\hbox{$\sim$}}\ }
\def\gl{\ \raise.5ex\hbox{$>$}\kern-.8em\lower.5ex\hbox{$<$}\ }
\def\roughly#1{\raise.3ex\hbox{$#1$\kern-.75em\lower1ex\hbox{$\sim$}}

}
%
%
%       This defines et al., i.e., e.g., cf., etc.

%

%
\def\npa#1#2#3{{Nucl. Phys. } A{#1} (#2) #3}
\def\npb#1#2#3{{Nucl. Phys. } B{#1} (#2) #3}
\def\pl#1#2#3{{Phys. Lett. } {#1}B (#2) #3}
\def\prl#1#2#3{{Phys. Rev. Lett. } {#1} (#2) #3}
\def\physrev#1#2#3{{Phys. Rev. } {#1} (#2) #3}

\relax

\def\dpart{\partial\kern .5ex\llap{\raise
1.7ex\hbox{$\leftrightarrow$}}\kern -.7ex {_\mu}}

\def\frac#1#2{{\textstyle{#1 \over #2}}}

\def\s2weak{\sin^2\theta_{\rm w}}

\def\({\left(}\def\){\right)}

\def\mayer{\vbox{\sl\centerline{Department of Physics, U.C.
San Diego}
\centerline{La Jolla, CA 92093-0319}}}

\def\Queens{\vbox{\sl\centerline{Department of Physics,
Stirling Hall}
\centerline{Queen's University, Kingston, Canada, K7L 3N6}}}
\def\Duke{\vbox{\baselineskip=14truept
\sl\centerline{Department of Physics}
\centerline{Duke University, Durham, NC 27708}}}

\def\[{\left[}
\def\]{\right]}
\def\({\left(}
\def\){\right)}

\noblackbox
\vskip .2in
\centerline{\titlefont Electromagnetic Moments of the}
\centerline{\titlefont Baryon Decuplet}
\bigskip
\centerline{Malcolm N. Butler\footnote{$^\dagger$}{Address as
of September  1, 1993:
Department of Astronomy and Physics, Saint Mary's University,
Halifax, NS, Canada B3H 3C3}}
\medskip
\Queens
\bigskip
\centerline{Martin J. Savage
\footnote{$^{\dagger\dagger}$}{ Address as of September 1, 1993:
Department of Physics, Carnegie Mellon University, Pittsburgh, PA,
USA 15213}
\footnote{$^*$}{SSC Fellow}}
\medskip
\mayer
\bigskip
\centerline{Roxanne P.\ Springer}
\medskip
\Duke
\medskip
\centerline{{\bf Abstract}}
We compute the leading contributions to the magnetic dipole and
electric quadrupole moments of the baryon decuplet in chiral
perturbation theory.  The measured value for the magnetic moment of the
$\Omega^-$ is used to determine the local counterterm for the magnetic
moments.  We compare the chiral perturbation theory predictions for the
magnetic moments of the decuplet with those of the baryon octet and
find reasonable agreement with the predictions of the large--$N_c$
limit of QCD.  The leading contribution to the quadrupole moment of the
$\Delta$ and other members of the decuplet comes from one--loop
graphs.   The pionic contribution is shown to be proportional to $I_z$
(and so will not contribute to the quadrupole moment of $I=0$ nuclei),
while the contribution from kaons has both isovector and isoscalar
components.  The chiral logarithmic enhancement of both pion and kaon
loops has a coefficient that vanishes in the $SU(6)$ limit.  The third
allowed moment, the magnetic octupole, is shown to be dominated by a
local counterterm with corrections arising at two loops.  We briefly
mention the strange counterparts of these moments.

\vfill
\line{UCSD/PTH 93-22, QUSTH-93-05, Duke-TH-93-56\hfill August 1993 }
\line{hep-ph/9308317 \hfil}
\eject

Static electromagnetic moments are a valuable tool for understanding
internal structure.  In nuclear physics, static moments played a
crucial role in understanding the strong tensor interaction arising
from one-pion exchange which lead to significant deviations from
spherical symmetry in simple nuclei like the deuteron.  In the same
way, hadronic structure can be investigated using static moments.  The
magnetic moments of the octet baryons have been understood in the
context of $SU(3)$ for many years \ref\cole{S.Coleman and S.L. Glashow,
\prl{6}{1961}{423}. }.  Leading, model independent corrections to these
$SU(3)$ relations have been computed in chiral perturbation theory
\ref\caldi{D.G. Caldi and H. Pagels, \physrev{D10}{1974}{3739}.
}\ref\jlms{ E. Jenkins {\it et al.}, \pl{302}{1993}{482}.} and have
been found to improve agreement with experimental data.

The electrostatic properties of the $\Delta$ and other members of the
baryon decuplet have received little theoretical attention since
experimental data has been scarce until recently.  The $\Omega^-$
magnetic moment was found to be $\mu_\Omega=-1.94\pm0.17\pm0.14\,\mu_N$
\ref\pdg{Particle Data Group, \physrev{D45}{1992}{1}.}, and a recent
measurement for the $\Delta^{++}$ moment using pion bremsstrahlung (a
model dependent extraction) found $\mu_{\Delta^{++}}=4.5\pm0.5\,\mu_N$
\ref\dpp{A. Bosshard {\it et al.}, \physrev{D44}{1991}{1962}.} (we note
that this result is not used by the PDG in their estimate \pdg ).  The
magnetic moments have been examined in the cloudy-bag model
\ref\cb{M.I. Krivoruchenko, Sov. J. Nucl. Phys., 45, (1987) 109.}, in
quark models \pdg\nref\nrqm{K.T. Chao,
\physrev{D41}{1990}{920}.}\nref\man{ H. Georgi and A. Manohar,
\pl{132}{1983}{183}.}--\ref\lfqm{ F. Schlumpf, SLAC-PUB-6218, (1993).},
in a Bethe-Salpeter model \ref\mitra{A.N. Mitra and A. Mittal,
\physrev{D29}{1984}{1399}.}, in the Skyrme model \ref\ed{G.S.Adkins,
C.R. Nappi and E. Witten, \npb{228}{1983}{552}.}\nref\skyrmea{ J.H.
Kim {\it et al.}, \npa{501}{1989}{835}.}--\ref\cohen{T.D. Cohen and W.
Broniowski, \physrev{D34}{1986}{3472}.}, using QCD sum rules
\ref\bely{V.M. Belyaev, CEBAF-TH-93-02, (1993).}, and also recently in
quenched lattice gauge theory \ref\latt{D.B. Leinweber, T. Draper and
R.M. Woloshyn, \physrev{D46}{1992}{3067}.}\ref\satoshi{S. Nozawa and
D.B. Leinweber, \physrev{D42}{1990}{3567}.}.  Our goal here is to
understand the magnetic moments in a model-independent, systematic way,
using chiral perturbation theory.

The decuplet baryons also have electric quadrupole and magnetic
octupole moments.  These moments have been studied recently using
quenched lattice QCD \latt\satoshi.  In chiral perturbation theory, the
pion one--loop graphs tend to dominate over either kaon one--loop
graphs or the local counterterm because of the presence of the
(calculable) chiral logarithm, $\log(M_\pi^2/\Lambda_\chi^2)$.  This
was seen in the calculation of the electric quadrupole matrix element
for the decay $\Delta \rightarrow N \gamma$ \ref\butlera{M.N.\ Butler,
M.J.\ Savage and R.P.\ Springer, Nucl.\ Phys.\ {\bf B399} (1993) 69.}
\ref\butlerb{M.N.\ Butler, M.J.\ Savage and R.P.\ Springer,
Phys.\ Lett.\ {\bf B304} (1993) 353; Erratum and Addendum, to be
published in Phys.\ Lett.\ {\bf B} (1993).}.  For the decuplet
quadrupole moments, however, we find that the pion one--loop
contributions are proportional to the third component of isospin, with
the result that baryons with $I_z=0$ receive no contribution from the
lowest order pion loop, and $I_z={1\over 2}$ baryons receive
approximately equal contributions from kaon and pion loops at lowest
order.  These are still formally dominant over the dimension six local
counterterm for the quadrupole moment.   We show that the octupole
moment is dominated by a dimension seven local counterterm with
corrections occurring first at two loops.

Heavy baryon chiral perturbation theory uses the chiral symmetry of QCD
to construct an effective low--energy theory to describe the dynamics
of the goldstone bosons associated with the spontaneous breaking of
chiral symmetry.  The baryons can be included in a consistent manner,
as shown in Ref.  \ref\heavybaryon{E.\ Jenkins and A.\ Manohar,
Phys.\ Lett.\ {\bf 255B} (1991) 558 ; Phys.\ Lett.\ {\bf 259B} (1991)
353.}.  For a review of the simplifications available in calculating
with this formalism see Ref.  \ref\jmhungary{E. Jenkins and A. Manohar,
Proceedings of the workshop on ``Effective Field Theories of the
Standard Model'', ed. U.  Meissner, World Scientific (1992)}.  The
decuplet of resonances as explicit degrees of freedom has been shown to
be important for most physical observables, and for consistency of the
perturbative expansion \heavybaryon \ref\heavybliz{E. Jenkins, Nucl.
Phys. {\bf B368} (1992) 190; Nucl.  Phys. {\bf B375} (1992) 561.}.
Further, Ref.  \ref\dasha{R. Dashen and A. Manohar, UCSD/PTH 93-16.}
shows that including the decuplet (in fact, the entire tower of $I=J$
baryons) is required in order for a low energy theory of pions and
nucleons to be unitary in the large--$N_c$ limit of QCD (where $N_c$ is
the number of colours).  The decuplet appears with couplings to the
pions satisfying a contracted $SU(2N_f)$ algebra (where $N_f$ is the
number of light flavours in the theory).  Ref. \ref\dashb{R. Dashen and
A.  Manohar, UCSD/PTH 93-18.} shows that the corrections to the
relations arising from this $SU(2N_f)$ symmetry occur first at order
$1/N_c^2$.

The leading SU(3) invariant local counterterm for the decuplet
magnetic moment is given by a dimension five operator \jlms\
\eqn\magct{L^{CT}_{M1} = -i {e\over M_N} \mu_c  q_k
\overline T{}^\mu_{v k} T^\nu_{vk} F_{\mu\nu}\>, }
where $T^\mu$ is the decuplet field and $q_k$ is the charge of the
$k$th baryon of the decuplet.  We have normalised the coefficient of
the operator so that the magnetic moment of the $k$th baryon is
$q_k\mu_c$ nuclear magnetons.  A simple tree-level fit to the magnetic
moment of the $\Omega^-$ hyperon gives $\mu_c=1.94\pm0.22\ \mu_N$
(where we have added the systematic and statistical errors of
$\mu_\Omega$ in quadrature).  The leading corrections to the magnetic
moments arise from the one-loop diagrams shown in \fig\moments{Leading
one--loop graphs contributing to the multipole moments of the decuplet
baryons.  The dashed lines correspond to charged goldstone bosons and
the wiggly line to photons.  $T$ is a decuplet baryon and $B$ is an
octet baryon.}.  A computation gives the following matrix elements:
\eqn\magttt{\CM^{TTT}= i{e\over 16\pi^2} \CH^2
(\overline T_v\cdot k T_{v\mu}-\overline T_{v\mu}T_v\cdot k)A^\mu
{2\over 3} \sum_i {\alpha^i \over f^2_{M_i}} \CF(\Delta m_i, M_i) \ \
\
 }
and
\eqn\magtbt{\CM^{TBT}=i{e\over 16\pi^2} \CC^2
(\overline T_v\cdot k T_{v\mu} - \overline T_{v\mu}T_v\cdot k)A^\mu
\sum_i {\beta^i \over f^2_{M_i}} \CF(\Delta m_i, M_i)\>,
}
where
\eqn\fdelm{\CF(\Delta m, M)= \Delta m \log\bigg( {M^2\over
\Lambda_\chi^2}\bigg)
+\sqrt{\Delta m^2-M^2} \log\bigg({\Delta m+\sqrt{\Delta
m^2-M^2+i\epsilon}\over
\Delta m-\sqrt{\Delta m^2-M^2+i\epsilon}}\bigg) \>,}
and $A^\mu$ is the electromagnetic gauge field.  The superscripts $TTT$
and $TBT$ denote the contribution from graphs with intermediate
decuplet and octet baryons respectively.  The mass splitting between
the external baryon and the baryon in the loop is $\Delta m_i$, the
mass of the relevant pseudogoldstone boson is $M_i$ ($i=\pi$ or $K$),
the chiral symmetry breaking scale is $\Lambda_\chi$, and the decay
constant of the meson in the loop is $f_{M_i}$ ($f_\pi=132$MeV and
$f_K=1.22f_\pi$).  The decuplet-octet-meson coupling constant is $\CC$
and the decuplet-decuplet-meson coupling constant is $\CH$.  The
constants $\alpha^i$ and $\beta^i$ are the product of the electric
charge of octet meson $i$ and SU(3) Clebsch--Gordan coefficients
(explicit values are given in the appendix).  The one--loop corrections
to the multipole moments depend only on the coupling constants $\cal C$
and $\cal H$.  Using ${\cal C}=-1.2\pm 0.1$, ${\cal H}=-2.2\pm0.6$
\butlera\ and the measured value of the $\Omega^-$ magnetic moment to
fix $\mu_c$, we predict the magnetic moments of the other members of
the baryon decuplet.  These results are shown in table 1 and also
graphically in \fig\mag{The magnetic moments of the decuplet baryons,
in units of nuclear magnetons.  The dark points are the moments derived
from the central values of $\CC$ and $\CH$ and the lighter lines are
the associated uncertainties.}.  It is clear from \mag\ that the SU(3)
violating corrections induced by the one-loop graphs (dominated by the
contribution from kaons) is small and that the tree level relation
(where the magnetic moment is proportional to the electric charge of
the baryon) is not badly broken.  The one--loop chiral perturbation
theory prediction for the $\Delta^{++}$ magnetic moment of
$\mu_{\Delta^{++}} = 4.0\pm 0.4$ (the tree-level result is
$\mu_{\Delta^{++}} = 5.8\pm 0.7$) agrees within errors with the recent
measurement (but model dependent extraction) of
$\mu_{\Delta^{++}}=4.5\pm0.5$ \dpp, and the results from  quenched
lattice QCD $\mu_{\Delta^{++}}=4.9\pm0.6$ \latt, but is significantly
smaller than the prediction of the naive quark model
$\mu_{\Delta^{++}}\sim 5.6$ \lfqm.  For the charged members of the
decuplet we agree with the quenched lattice computations \latt\  but
differ in the predictions for the magnetic moments of the neutral
baryons. Note that the neutral baryon magnetic moments do not depend on
the local leading counterterm that appears for the charged baryons,
making these predictions independent of the measured value of
$\mu_{\Omega^-}$.

As mentioned earlier, the axial matrix elements in $I=J$ baryons such
as $N$ and $\Delta$ must obey a contracted $SU(2N_f)$ algebra in the
large $N_c$ limit of QCD \dasha .  This results from the need for the
low energy theory of baryons and goldstone bosons to be unitary in the
large $N_c$ limit.  Further, it was shown that this requires the
$1/N_c$ correction to the axial matrix elements be proportional to the
leading term.  Therefore, relationships between axial matrix elements
have vanishing $1/N_c$ corrections, but are corrected at order
$1/N_c^2$ and higher \dashb.  A similar argument can be constructed for
the matrix elements of the isovector magnetic moment operator.  We
expect that they satisfy the relations of a contracted $SU(2N_f)$
algebra up to corrections arising from terms $1/N_c^2$ and higher in
the $1/N_c$ expansion.  In this limit the isovector magnetic moments
satisfy
\eqn\large{  {\mu_{\Delta^{++}}-\mu_{\Delta^-}\over \mu_p-\mu_n} =
{9\over 5} + {\cal O}({1\over N_c^2})}
and
\eqn\largeb{  {\mu_{\Delta^+}-\mu_{\Delta^0}\over \mu_p-\mu_n} =
{3\over 5} + {\cal O}({1\over N_c^2})\ \ \  .}
Our analysis of magnetic moments is a non-trivial test of these
relations.  The local counterterm given in  \magct\ has both isoscalar
and isovector components, since it is proportional to the electric
charge operator.  At tree--level
\eqn\treen{  \left(
{\mu_{\Delta^{++}}-\mu_{\Delta^-}\over\mu_p-\mu_n}
\right)_{\rm tree} =
-{3\mu_{\Omega}\over \mu_p-\mu_n} \sim 1.2 }
and
\eqn\treenb{ \left( {\mu_{\Delta^+}-\mu_{\Delta^0}\over\mu_p-\mu_n}
\right)_{\rm tree} =
-{\mu_{\Omega}\over\mu_p-\mu_n} \sim 0.4 \ \ \ ,}
which are about $2/3$ the values expected in the  large $N_c$ limit.
Including the one--loop graphs improves the situation  somewhat and we
find that
\eqn\loopn{ \left( {\mu_{\Delta^{++}}-\mu_{\Delta^-}\over\mu_p-\mu_n}
\right)_{\rm one-loop} = 1.35\pm 0.15\ \ \,}
and
\eqn\loopnb{ \left( {\mu_{\Delta^+}-\mu_{\Delta^0}\over \mu_p-\mu_n}
\right)_{\rm one-loop} = 0.45\pm 0.05 \ \ .}
Despite the fact that both quantities are still smaller than the
numbers expected from large-$N_c$ QCD, the one-loop corrections tend to
reduce the discrepancy in each case.  There are modifications to the
large $N_c$ relations from terms subleading in the $1/N_c$ expansion
and also corrections at the $25\%$ level from terms higher order in the
chiral expansion that may improve the agreement.

The quadrupole moment for each of the decuplet baryons receives a
contribution from both long--distance physics in the form of pion and
kaon loops, and from short distance physics in the form of a local
counterterm with an unknown coefficient.  This dimension six
counterterm has the form
\eqn\qct{\CL_{E2}^{CT}= Q_{CT} {e\over \Lambda_\chi^2}
q_i(\overline T_{vi}^\mu T_{vi}^\nu +\overline T_{vi}^\nu T_{vi}^\mu
-{1\over 2} g^{\mu\nu}
\overline T_{vi}^\sigma T_{vi\sigma})v^\alpha \partial_\mu
F_{\nu\alpha}  \>.}
The contribution to the quadrupole moment from the diagrams in
\moments\ are formally enhanced over the naive contribution from the
local counterterm by a chiral logarithm, $\log(M^2/\Lambda_\chi^2)$,
and we will neglect the contributions from the local counterterm,
taking $Q_{CT}\sim 0$ for the rest of this discussion.  The explicit
contributions from the graphs in \moments\  are
\eqn\qttt{ Q^{TTT}=-i {e\over 16\pi^2} {2\over 9}\CH^2\omega
(\overline T_v\cdot k T_{v\mu}+\overline T_{v\mu} T_v\cdot
k-{1\over 2}k_\mu\overline T_v\cdot T_v)A^\mu
\sum_i {\alpha^i\over f^2_{M_i}}\CG(\Delta m_i, M_i) }
and
\eqn\qtbt{ Q^{TBT}= i {e\over 16\pi^2} \CC^2
{\omega\over 6}
(\overline T_v\cdot k T_{v\mu}+\overline T_{v\mu} T_v\cdot
k-{1\over 2}k_\mu\overline T_v\cdot T_v) A^\mu
\sum_i {\beta^i\over f^2_{M_i}}\CG (\Delta m_i, M_i)
\>,}
where
\eqn\func{\CG (\Delta m, M) = \log\bigg({M^2\over
\Lambda_\chi^2}\bigg)
+{\Delta m\over\sqrt{\Delta m^2-M^2}}
\log\bigg({\Delta m+\sqrt{\Delta m^2-M^2+i\epsilon}
     \over \Delta m-\sqrt{\Delta m^2-M^2+i\epsilon}}\bigg)
\ \ \ .}
As before, $\Delta m_i$ is the mass splitting between the external and
loop baryon, $M_i$ is the mass of the goldstone boson in the loop, and
$f_{M_i}$ is the meson decay constant.  The notation for $\alpha$ and
$\beta$ is the same as for the magnetic moment equations.  We can
extract quadrupole moments from this calculation by using the
definition of the quadrupole interaction energy,
\eqn\hquad{H^Q=-{1\over 6} \sum_{ij} Q_{ij} {\partial E_i\over
\partial x^j}
\>,}
where $E$ is the electric field and $Q_{ij}$ is the quadrupole tensor
which is symmetric and traceless.  The quadrupole moment is defined to
be $Q_{zz}$, and can be extracted from \qttt\ and \qtbt.  The results
for the various members of the decuplet, neglecting the formally
subdominant counterterm of \qct, are shown in Table 2 and graphically
in \fig\quad{The quadrupole moments of of the decuplet baryons.  The
contribution from the local counterterm is subleading and we have set
it to zero.  The quadrupole moments here come from one-loop graphs
only. The dark points are the moments derived from the central values
of $\CC$ and $\CH$ and the lighter lines are the associated
uncertainties.}.  These moments are large, and comparable to the
moments of light nuclei such as the deuteron ($Q_D=2.8\times 10^{-27}
e-{\rm cm}^2$).  With such large moments, the presence of constituent
$\Delta$'s in nuclei might have a significant effect on nuclear
quadrupole moments.  Naively, the pion loop graphs should be
logarithmically enhanced over the kaon loop graphs.  Yet, the
Clebsch-Gordan coefficients (given in the appendix) are such that the
quadrupole moment generated by the pion loops depend only upon the
$I_z$ quantum number of the baryon (the kaon loops have both isovector
and isoscalar dependence).    This is distinctly different from the
dependence of the local counterterm which depends on the charge of the
baryon.  The importance of this result becomes apparent when
considering the quadrupole moment of a nucleus, in particular an $I=0$
nucleus such as the deuteron.  One might imagine that the intrinsic
quadrupole moment of the $\Delta$ would contribute to the quadrupole
moment of a nucleus through virtual $\Delta$ states.  However, as most
of the intrinsic quadrupole moment of the $\Delta$ depends on $I_z$,
this contribution to the quadrupole moment of an $I=0$ nucleus
vanishes.  Hence, the $\Delta$ contribution to the quadrupole moment of
the deuteron is greatly suppressed over naive expectations, appearing
first from the kaon loop contribution.

Our values for the quadrupole
moments are not always consistent with the values found in quenched
lattice computations \latt, though all but the neutral baryons agree
within errors.  In particular, where we find that the dominant
component behaves as $I_z$, lattice computations find behaviour more
consistent with dependence upon the baryon charge.

Another interesting, perhaps more mysterious result that can be found
by examining the Clebsch-Gordan coefficients in the appendix is that
the coefficient of both the $\log(M_\pi^2/\Lambda_\chi^2)$ and
$\log(M_K^2/\Lambda_\chi^2)$ terms are proportional to ${4\over 9}\CH^2
- \CC^2$. This vanishes when $\CH/\CC = 3/2$, which is exactly the
relationship satisfied in the $SU(6)$ limit.  In this $SU(6)$ limit,
the contribution to the quadrupole moment from these one-loop graphs
arises entirely from the mass splittings amongst the baryons.  We can
reconcile our results with that of quenched lattice QCD if indeed the
axial couplings are very close to their $SU(6)$ values.  The quadrupole
moments would then receive a non-negligible, and possibly dominant,
contribution from the incalculable local counterterm (which we have
neglected for our discussions), giving the characteristic dependence on
the baryon charge that the lattice calculations find.  Our central
value predictions would then be substantially smaller in magnitude than
those obtained using the experimentally fit values of $\CH$ and $\CC$.

In dealing with the magnetic moments of the decuplet, we saw that the
large $N_c$ limit of QCD gave results consistent with those of chiral
perturbation theory calculations.  For the quadrupole moments, the
large $N_c$ limit of the one-loop contribution approaches a constant
value.  This is because the relationships between axial coupling
constants $\CF$, $\CD$, $\CC$, and $\CH$ approach their $SU(6)$ values
\dasha\dashb, and the hyperfine mass splittings between the baryons
vanish as $1/N_c$ \ref\liz{E. Jenkins UCSD/PTH 93-19 (1993).}.  The
coefficient of the quadrupole counterterm and the $1/N_c^2$ corrections
to the hyperfine mass splittings are needed in order to make a more
explicit comparison between the large $N_c$ predictions and chiral
perturbation theory results for the quadrupole moments.

Finally, the decuplet baryons could also have a magnetic octupole
moment.  We can construct a dimension seven local counterterm for
this moment, of the form
\eqn\oct{\CL^{CT}_{M3}= e{\Theta\over \Lambda_\chi^3}q_i(\overline
T^\mu_{vi}
S^\nu_v T^\alpha_{vi} +\overline T^\nu_{vi} S^\alpha_v T^\mu_{vi}
+\overline T^\alpha_{vi} S^\mu_v T^\nu_{vi})
\epsilon_{\alpha\beta\lambda\sigma}v^\beta \partial_\mu\partial_\nu
F^{\lambda\sigma}\>,}
where $\Theta$ is an unknown coefficient.  This tensor structure, in
particular the three derivatives of the electromagnetic field, does
not appear in the one--loop graphs shown in \moments.  Therefore, the
magnetic octupole moment will be dominated by the local counterterm
and corrections can first occur from two--loop diagrams.

In addition to the electrostatic moments of these baryons we can
examine their strange moments.  Strange moments of the nucleons as
suggested in \ref\dav{D.B. Kaplan and A. Manohar,
\npb{310}{1988}{527}.} have been the subject of an immense amount of
both theoretical and experimental interest.  Estimates of the size of
these moments have been made for the octet baryons in the context of
different hadronic schemes \ref\hen{W. Koepf, E.M. Henley and S.J.
Pollock, \pl{288}{1992}{11}.} \ref\mus{M.J. Musolf and M. Burkardt,
CEBAF TH-93-01.}.  The strange moments of the decuplet baryons may
never be measured, yet we are able to see what form they will have in
the language of chiral perturbation theory.  The strange magnetic and
quadrupole moments could be substantially different from their
electromagnetic counterparts.  Since the strange charge operator has
both flavour octet and singlet components, there are two unknown
counterterms for each strange moment, with $SU(3)$ structure
\eqn\strange{{\cal L}\sim
S\ \overline{T}^{abc}Q^{(s)d}_cT_{abd}\ \ +\ \  \sigma\
\overline{T}^{abc}T_{abc}Q^{(s)\alpha}_\alpha\ \ \ ,}
where the strange charge matrix is $Q^{(s)}$=diag(0,0,1) and $S$ and
$\sigma$ are unknown coefficients.  For investigating the baryon sea,
however, we are most interested in looking at the strange moments of
the non-strange baryons, namely the $\Delta$'s.  The first term, $S$,
does not contribute, which leaves one unknown counterterm, $\sigma$,
that contributes equally to all baryons in the decuplet, yet is unclear
how to determine experimentally.  We can compute corrections to the
strange magnetic moments and also the dominant contribution to the
strange quadrupole moment just as we did in the electrostatic case.
For these strange moments the pion loops do not contribute (they do not
carry strange charge) but both charged and neutral kaons will
contribute.  Therefore, we expect the strange quadrupole moment to be
much smaller than the electrostatic counterpart for large $I_z$
baryons, with the other quadrupole moments perhaps comparable in size
to the electrostatic ones.  The strange magnetic moment may be the same
size as the electrostatic magnetic moment.  Since the strange charge is
an isoscalar, the moments of each of the $\Delta$'s are identical.
The one--loop induced quadrupole moments are proportional to ${4\over
9}\CH^2-\CC^2$ (up to isospin breaking mass differences), a quantity
that vanishes in the $SU(6)$ limit.  (Unlike the case for the
electromagnetic quadrupole moment, the intermediate baryons
contributing to these quadrupole moments are all isospin degenerate.)
If the axial couplings are near the $SU(6)$ point, as there is strong
evidence to suggest, then the strange moments of the $\Delta$'s are
each dominated by one incalculable local counterterm.  These quantities
are of theoretical interest as the appearance of possibly another
non-zero strange matrix element in non-strange hadrons.

In conclusion, we have discussed the electrostatic properties of the
$\Delta$ and other members of the baryon decuplet.   Using chiral
perturbation theory we have computed the leading non-analytic
contributions to the magnetic dipole and electric quadrupole moments,
and shown that the leading contribution to the octupole moment is from
a local dimension seven counterterm with corrections arising at
two-loops.  We have compared our prediction for the $\Delta^{++}$
magnetic moment with its recent model dependent extraction from pion
bremsstrahlung data and found it to be in good agreement.  The one-loop
computation moves the isovector magnetic moments into better agreement
with the predictions of large-$N_c$ QCD compared to the tree-level
results.  Although the quadrupole moments of the decuplet have not been
measured yet, there may be some hope for such measurements at CEBAF.
We computed the leading contribution to the quadrupole moments from
long-distance pion and kaon loops, which are formally dominant over the
dimension six local counterterm.  The pion contribution depends only on
$I_z$ and hence the contribution from the intrinsic quadrupole moment
of $\Delta$'s to that of an $I=0$ nucleus from the $\Delta$ components
in the nuclear wavefunction is suppressed.   This is an important
result particularly for the deuteron since the magnitude of the
$\Delta$ quadrupole moments are comparable to that of the deuteron.
Further, the formally dominant terms of the form
$\log(M_\pi^2/\Lambda_\chi^2)$ and $\log(M_K^2/\Lambda_\chi^2)$ vanish
when the axial couplings approach their $SU(6)$ limit.

We have compared our results to those obtained in quenched lattice QCD
\latt\ and find that the magnetic moments of the charged baryons agree
well.  This is not unexpected since they are dominated by the local
counterterm that is fixed experimentally.  This agreement does not
exist for the neutral baryons, which have no counterterm dependence.
The predictions for the charged baryon electric quadrupole moments also
agree within errors, yet have a different dependence on the baryon
isospin.  Again, there is not agreement for the neutral baryons.  Our
leading contribution for the quadrupole moment arises from pion loops
and depends on $I_z$ only.  The lattice computation indicates that the
quadrupole moment depends on the charge of the baryon.  These two
results can be reconciled if the axial coupling constants $\CC$ and
$\CH$ satisfy $SU(6)$ relations as required in the large $N_c$ limit
and also approximately found experimentally \butlera\jmhungary.  In
this scenario the quadrupole moments receive a non-negligible and
potentially dominant contribution from the local (incalculable)
counterterm.

We expect that some of our predictions will be tested at CEBAF, and
that measurements of the quadrupole moments in particular may help test
the validity of the heavy baryon chiral perturbation theory approach in
understanding low--energy QCD.  In addition, we expect the
comparison to help determine if the contracted $SU(2N_f)$ algebra is a
useful symmetry for describing low--energy  hadronic properties.

\bigskip\bigskip

MNB acknowledges the support of the Natural Science and Engineering
Research Council  (NSERC) of Canada.  MJS acknowledges the support of a
Superconducting Supercollider National Fellowship from the Texas
National Research Laboratory Commission under grant  FCFY9219, and the
hospitality of the Aspen Institute for Physics where much of this work
was done. RPS acknowledges the support of DOE grant DE-FG05-90ER40592.

\vfil\eject

\noindent {\bf Appendix}

SU(3) Clebsch-Gordan coefficients $\alpha$ and $\beta$ used for
the evaluation of decuplet electromagnetic moments.
For brevity, only the intermediate baryon state is given as a
label. The boson index is implicit.

{$\Delta^{++}$}
$$
\hss\vbox{\halign{$#$\hfil\tabskip=10em&$#$\hfil\cr
\alpha_{\Delta^+}={1\over 3}&
\beta_{p}=1\cr
\alpha_{\Sigma^{*+}}={1\over 3}&
\beta_{\Sigma^+}=1\cr}}\hss
$$

{$\Delta^+$}
$$
\hss\vbox{\halign{$#$\hfil\tabskip=10em&$#$\hfil\cr
\alpha_{\Delta^0}={4\over 9}&
\beta_{n}={1\over3}\cr
\alpha_{\Delta^{++}}=-{1\over 3}&\cr
\alpha_{\Sigma^{*0}}={2\over 9}&
\beta_{\Sigma^0}={2\over3}\cr}}\hss
$$

{$\Delta^0$}
$$
\hss\vbox{\halign{$#$\hfil\tabskip=10em&$#$\hfil\cr
\alpha_{\Delta^-}={1\over 3}&\cr
\alpha_{\Delta^+}=-{4\over 9}&
\beta_{p}=-{1\over3}\cr
\alpha_{\Sigma^{*-}}={1\over 9}&
\beta_{\Sigma^-}= {1\over3}\cr}}\hss
$$

{$\Delta^-$}
$$
\hss\vbox{\halign{$#$\hfil\tabskip=10em&$#$\hfil\cr
\alpha_{\Delta^0}=-{1\over 3}&
\beta_{n}=-1\cr}}\hss
$$

{$\Sigma^{*+}$}
$$
\hss\vbox{\halign{$#$\hfil\tabskip=10em&$#$\hfil\cr
\alpha_{\Sigma^{*0}}={2\over 9}&
\beta_{\Sigma^0}={1\over6}\cr
&\beta_{\Lambda}={1\over2}\cr
\alpha_{\Xi^{*0}}={4\over 9}&
\beta_{\Xi^0}={1\over3}\cr
\alpha_{\Delta^{++}}=-{1\over 3}&\cr}}\hss
$$

{$\Sigma^{*0}$}
$$
\hss\vbox{\halign{$#$\hfil\tabskip=10em&$#$\hfil\cr
\alpha_{\Sigma^{*-}}={2\over 9}&
\beta_{\Sigma^-}={1\over6}\cr
\alpha_{\Sigma^{*+}}=-{2\over 9}&
\beta_{\Sigma^+}=-{1\over6}\cr
\alpha_{\Xi^{*-}}={2\over 9}&
\beta_{\Xi^-}={1\over6}\cr
\alpha_{\Delta^{+}}=-{2\over 9}&
\beta_{p}=-{1\over6}\cr}}\hss
$$

{$\Sigma^{*-}$}
$$
\hss\vbox{\halign{$#$\hfil\tabskip=10em&$#$\hfil\cr
\alpha_{\Sigma^{*0}}=-{2\over 9}&
\beta_{\Sigma^0}=-{1\over6}\cr
&\beta_{\Lambda}=-{1\over2}\cr
\alpha_{\Delta^{0}}=-{1\over 9}&
\beta_{n}=-{1\over3}\cr}}\hss
$$

{$\Xi^{*0}$}
$$
\hss\vbox{\halign{$#$\hfil\tabskip=10em&$#$\hfil\cr
\alpha_{\Xi^{*-}}={1\over 9}&
\beta_{\Xi^-}={1\over3}\cr
\alpha_{\Omega^{-}}={1\over 3}&\cr
\alpha_{\Sigma^{*+}}=-{4\over 9}&
\beta_{\Sigma^+}=-{1\over3}\cr}}\hss
$$

{$\Xi^{*-}$}
$$
\hss\vbox{\halign{$#$\hfil\tabskip=10em&$#$\hfil\cr
\alpha_{\Xi^{*0}}=-{1\over 9}&
\beta_{\Xi^0}=-{1\over3}\cr
&\beta_{\Lambda}=-{1\over2}\cr
\alpha_{\Sigma^{*0}}=-{2\over 9}&
\beta_{\Sigma^0}=-{1\over6}\cr}}\hss
$$

{$\Omega^-$}
$$
\hss\vbox{\halign{$#$\hfil\tabskip=10em&$#$\hfil\cr
\alpha_{\Xi^{*0}}=-{1\over3}&
\beta_{\Xi^0}=-1\cr}}\hss
$$

\listrefs
\vfil\eject
\centerline{\bf Table 1}
\midinsert\narrower
Magnetic moments of the baryon decuplet from chiral
perturbation theory.  Uncertainties reflect uncertainties
in the couplings $\CC$ and $\CH$, and in the magnetic moment
of the $\Omega^-$ used to constrain the local counterterm.
\endinsert

$$
\hss\vbox{\halign{#\hfil\tabskip=4em&\hfil#&\vrule#&\strut#\hfil&\hfil

#\cr
&\omit\hfil$\mu\ (\mu_N)$\hfil&&&\omit\hfil$\mu\ (\mu_N)$\hfil\cr
\noalign{\hrule}
$\Delta^{++}$&$4.0\pm0.4$&&$\Sigma^{*+}$&$2.0\pm0.2$\cr
$\Delta^+$&$2.1\pm0.2$&&$\Sigma^{*0}$&$-0.07\pm0.02$\cr
$\Delta^0$&$-0.17\pm0.04$&&$\Sigma^{*-}$&$-2.2\pm0.2$\cr
$\Delta^-$&$-2.25\pm0.25$&&&\cr
&&&$\Xi^{*0}$&$0.10\pm0.04$\cr
$\Omega^-$\ \ (input)&$-1.94\pm0.22$&&$\Xi^{*-}$&$-2.0\pm0.2$\cr
}}\hss
$$

\vskip 1truein

\centerline{\bf Table 2}
\midinsert\narrower
Quadrupole moments of the baryon decuplet arising from one--loop
graphs in chiral
perturbation theory.  Uncertainties reflect uncertainties
in the couplings $\CC$ and $\CH$.
\endinsert

$$
\hss\vbox{\halign{#\hfil\tabskip=4em&\hfil#&\vrule#&\strut#\hfil&\hfil
#\cr
&\omit\hfil$Q\ (10^{-27}e-{\rm cm}^2)$\hfil&&&\omit\hfil$Q\
(10^{-27}e-{\rm cm}^2)$\hfil\cr
\noalign{\hrule}
$\Delta^{++}$&$-0.8\pm0.5$&&$\Sigma^{*+}$&$-0.7\pm0.3$\cr
$\Delta^+$&$-0.3\pm0.2$&&$\Sigma^{*0}$&$-0.13\pm0.07$\cr
$\Delta^0$&$0.12\pm0.05$&&$\Sigma^{*-}$&$0.4\pm0.2$\cr
$\Delta^-$&$0.6\pm0.3$&&&\cr
&&&$\Xi^{*0}$&$-0.35\pm0.2$\cr
$\Omega^-$&$0.09\pm0.05$&&$\Xi^{*-}$&$0.2\pm0.1$\cr
}}\hss
$$

\listfigs
\end